\documentclass{elsart}

\usepackage{graphicx}

\usepackage[hang,center]{subfigure}

\usepackage{amssymb}
\begin{document}

\begin{frontmatter}
\title{Methods for measuring pedestrian density, flow, speed and 
direction with minimal scatter}

\author[inst1]{B. Steffen}
\author[inst1]{A. Seyfried}
\address[inst1]{J\"ulich Supercomputing Centre (JSC), 
Forschungszentrum J\"ulich GmbH,\\ 52425 J\"ulich, Germany}

\begin{abstract}
The progress of image processing during recent years allows the  measurement
of pedestrian characteristics on a ``microscopic'' scale with low
costs. However, density and flow are concepts of fluid mechanics defined for
the limit of infinitely many particles. Standard methods of measuring these
quantities locally (e.g. counting heads within a rectangle) suffer from large
data scatter. The remedy of averaging over large spaces or long times reduces  
the possible resolution and inhibits the gain obtained by the new
technologies. 

In this contribution we introduce a concept for measuring microscopic
characteristics on the basis of pedestrian trajectories. Assigning a personal
space to every pedestrian via a Voronoi diagram reduces the density
scatter. Similarly, calculating direction and speed from position differences
between times with identical phases of movement gives low-scatter sequences
for speed and direction. Closing we discuss the methods to obtain reliable
values for derived quantities and new possibilities of in depth analysis of
experiments. The resolution obtained indicates the limits of stationary state
theory. 
\end{abstract}

\begin{keyword}
Video tracking, Voronoi diagram, pedestrian modeling, velocity measurement,
pedestrian density
\end{keyword}
\end{frontmatter}

\section{Introduction}

For the design of pedestrian facilities concerning safety and level of service
\cite{Fruin1971,Predtechenskii1978,Weidmann1993,HCM,DiNenno2002}, pedestrian
streams are usually characterized by basic quantities like density or flow
borrowed from fluid dynamics. Up to now the experimental data base is
insufficient and contradictory
\cite{Weidmann1993,Schadschneider,Seyfried2007d} and thus asks for additional
experimental studies as well as improved measurement methods. Most
experimental studies of pedestrian dynamics use the classical definition of 
the density in an area by $D=\frac{N}{|A|}$, where $N$ gives the number of
pedestrians in the area $A$ of size $|A| \, [m^2]$; see
e.g. \cite{Togawa1955,Hankin1958,Oeding1963,Navin1969,Tanaboriboon1986,Mori1987,Lam2003,Hoogendoorn2005,Rupprecht2007,Fang2008}. 

An obvious problem is that this gives an averaged density for a specific area,
not a density distribution $p(\vec{x})  \, [1/m^2]$, and passing to the limit
of areas of size zero is obviously not a well defined procedure. Also the
choice of the geometry of $A$ is important. For large convex areas it can be
expected that finite size and boundary effects as well as influences of the
shape of the area can be neglected, though it is almost always possible to cut
out fairly large areas (of complicated shape) containing no person. Further,
the design of pedestrian facilities is usually restricted to an order of
magnitude of $1$ to $10m$ and rectangular geometries. In addition the number
of pedestrians $N$ is small and thus the scatter of local measurements has the
same order of magnitude as the quantity itself, see e.g. the time development
of the density in front of a bottleneck in \cite{Seyfried2007c}. Often, due to
cost restrictions the density was measured at a certain point in time, while
the measurement of speed was averaged over a certain time interval
\cite{Hankin1958,Navin1969}. But the process of measurement and averaging
influences the resulting data even for systems with few degrees of freedom as 
the movement along a line, see \cite{Seyfried2008,Seyfried2009a}. 

The large progress in video techniques during recent years has made feasible
the gathering of much more detailed data on pedestrian behavior, both in
experiments and in real life situations, than was possible only a decade ago
\cite{Hoogendoorn2005,Johansson2008,Boltes}. The higher detail asks for a
reevaluation of the methods of defining and measuring basic quantities like
density, flow and speed, as the methods to get time and space averages
encompassing a hundred persons over minutes may not be suitable for a
resolution of a second and a square meter. Basic quantities of pedestrian
dynamics are the density $D\;[1/m^2]$ in an area A, the the velocity $\vec{v}
\; [m/s]$ and speed $s = \|\vec{v}\|$ of a person or a group of persons, and the
flow through a door or across a specific line $F\,[1/s]$. Measurements give
averages of these quantities, and trying to carry the measurement to the
infinitesimal limit is obviously not reasonable. Below we indicate where the 
limits for the possible resolution are, and give methods that allow to go
fairly close to these limits. The methods presented here are based on video
tracking of the head from above, but tracking of e.g. a shoulder or the chest
might do even better, though they are more difficult to obtain. 

As density is used as one (important) parameter in modeling peoples behavior,
a perceived density would be best, but the details of perception are
completely unknown. However, the perception definitely does not show the rapid
fluctuations in time that the classical measurements do. 

There is not a `right' definition for density, flow or velocity  of
pedestrians (though some definitions may be wrong), the definitions treated
here are more or less useful for a certain purpose and allow more or less
resolution.  

All the examples in the paper are from experiments conducted in J\"ulich and 
D\"usseldorf and described in \cite{Boltes} and \cite{Seyfried2009a}.

\section{Direct Measurements} 
\subsection{Measuring density} 

In pedestrian dynamics, density is persons per area. The density at a point is
a mathematical abstraction, as persons are discrete and extended in space, and
the density is well defined only on a scale large enough to be able to ignore
discreteness and small enough to be homogeneous. These two conditions are no
problem in fluid mechanics with $ > 10^{18}$ particles per $mm^3$, but in
treating  pedestrians they are conflicting. The definition is directly
reflected by the standard procedure of measuring it. Here a certain area,
usually a rectangle, is laid out, and the number of persons inside this area
is counted. With $N$ pedestrians in the measurement area $A$ the standard
definition is 

\begin{equation}
D_s = \frac{N}{|A|}.
\label{DSTAND}
\end{equation}

This method has two drawbacks. The less important one is that occasionally
``in'' and ``out'' has to be assigned arbitrarily, usually by head
position. More important, the density depends discontinuously on time and on
the exact placement of the measuring area, with large jumps for small
area. These jumps can be reduced by taking averages over time and/or position
of the area, at the cost of resolution.

A way to define point values of the density is having every person $i$ produce
a density distribution $p_i(\vec{x})$ - a non-negative function with unit
integral $\int p_i(\vec{x})\; d\vec{x} = 1$. A step function
(e.g. $p_i(\vec{x}) = 1 /(2 r^2 \pi)  \, \mbox{for} \, \|\vec{x} - \vec{x}_i\| <
r $) , a linear function of the distance $p_i(\vec{x}) = max(0,h \, (r - \|\vec{x} -
\vec{x}_i\|)  $, or a Gaussian may do for different purposes, e.g. in molecular
dynamics. The central problem is the extention $r$ (or $\sigma$  for Gaussian)
of this function. The requirements are that in a situation  considered
homogeneous, the resulting density from a group of people will not  show too
much variations, while inhomogeneities will not be masked by too large a width
of the individual density function. These requirements can't be met with a
fixed extention $r$ (or  $\sigma$) of $p_i$. Much more involved but no problem
for computers, is the following adaptive procedure for step functions:

\begin{itemize}
\item With a given set of trajectories of $M$ persons
  $\{\vec{x}_1(t),\vec{x}_2(t), ..., \vec{x}_{M}(t)\}$ in two dimensions
  $\vec{x}=(x,y)$ assign an exact position $\vec{x}_i(t_0)$ at time $t_0$ for
  every person $i$.        
\item Compute the Voronoi-diagram \cite{V-D} for these positions, giving 
cells $A_i$ for each person $i$. 
\item Compute the size of the cell ${|A_i|}\,= {\int_{A_i} d\vec{x}}$ and define 
the density distribution for all persons: 

\begin{equation}
 p_i(\vec{x}) =  \left \{\begin{array}{r@{\quad:\quad}l}
\frac{1}{|A_i|} & \vec{x} \in A_i \\
0 & \mbox{otherwise} 
\end{array} \right \} \qquad \mbox{and}  \qquad  p(\vec{x}) = \sum_i p_i(\vec{x}) .
\end{equation}

\end{itemize}

\begin{figure}[thb]
  \begin{center}
    \includegraphics[width=5.cm,angle = -90]{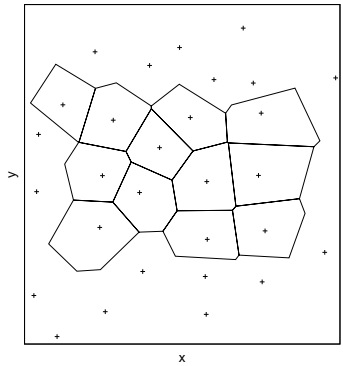}
\caption{Positions of persons jamming in front of an exit with exemplary
  Voronoi cells}  \label{VDplot}  
  \end{center}
\end{figure}

The Voronoi diagram Fig.\ref{VDplot} assigns to every point the area closer to
this point than to any other one. It has been used for defining density in the
fields of granular media \cite{Song2008}, molecular physics \cite{Guer} and
neuroanatomy \cite{Duy}, and finds further widespread applications in graphics
and mesh generation for PDEs. Here the program VRONI is used
\cite{Held2001}. 

Problems arise for the persons at the rim of a group, for which the Voronoi
cell may extend to infinity. Restricting the individual cells in size helps,
but the details of this are somewhat arbitrary. For our procedure to measure
the density in the given area, we used a restriction to $2 m^2$, which is
active only for a few cells. In the presence of walls, the Voronoi cell is
restricted to the area inside. It has been proposed by Pauls \cite{Pauls1980}
for stairs to use virtual walls 15 cm inside the  physical walls, but this
distance depends on details of the walls like roughness etc. and may be
different for level floors.

With the procedure defined above different density definitions are possible. 
To define a density in an area $A$ we choose 

\begin{equation}
  D_V = \frac{\int_A p(\vec{x}) \, d\vec{x}}{|A|}
\label{DVORONI} 
\end{equation}

Another Voronoi diagram based definition which has some merits with respect to
the observation area required is

\begin{equation}
  D_{V'} = \frac{N}{\sum_{i=1}^{N} |A_i|}
\label{DVORONISTR} 
\end{equation}

In this definition only the sizes of the Voronoi cells for the persons inside
$A$ are used. Both $D_V$ and $ D_{V'}$ are not very sensitive to the
positioning of the area of measurement. $D_V$ is preferable for small
measurement areas inside a crowd, where $D_{V'}$ may not be defined if there
is no person inside the area. On the other hand, $D_V$ requires useful Voronoi
cells for all people whose cell overlaps with the area A, not just for those
whose head is inside A as $D_{V'}$ does. Therefore it is more likely to be
affected by proximity to the rim of a group. 

\begin{figure}[thb]
\begin{center}
    \subfigure[Classical density $D_s$]{\label{VRON:xsuba} 
      \includegraphics[width=4.3cm,angle = 0]{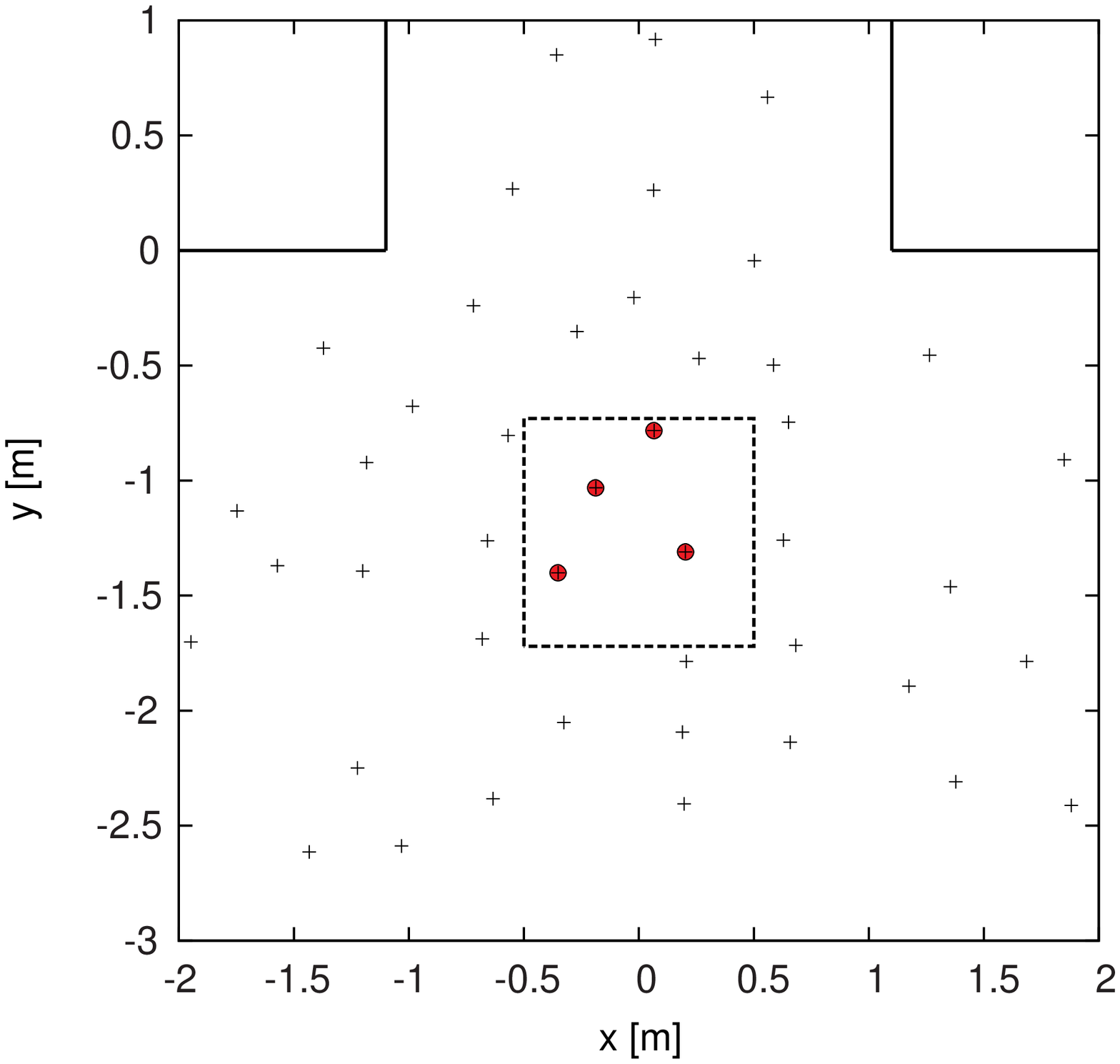}}
    \subfigure[Voronoi density $D_{V'}$]{\label{VRON:xsubb}
    \includegraphics[width=4.3cm,angle = 0]{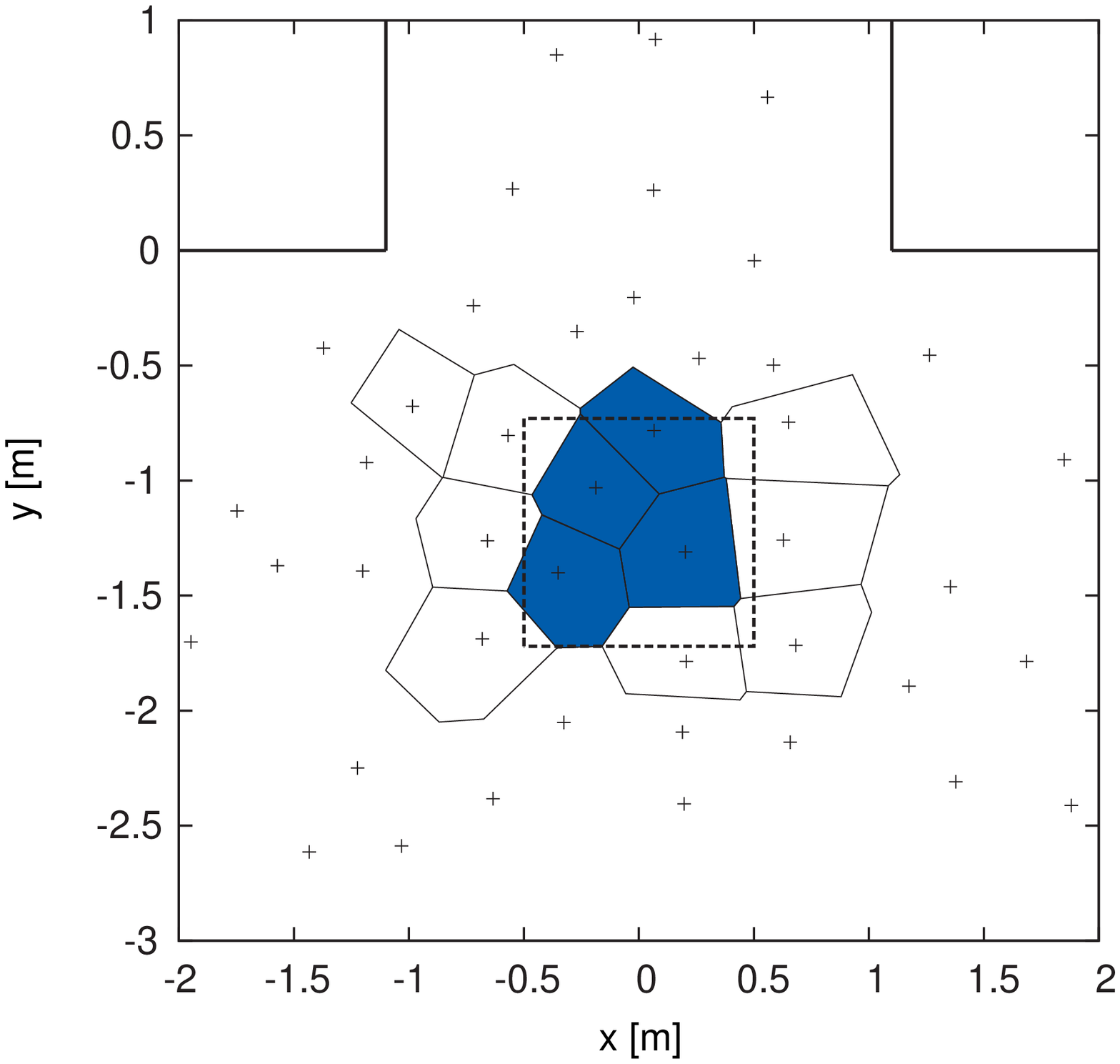}}
    \subfigure[Voronoi density $D_{V}$]{\label{VRON:xsubc} 
      \includegraphics[width=4.3cm,angle = 0]{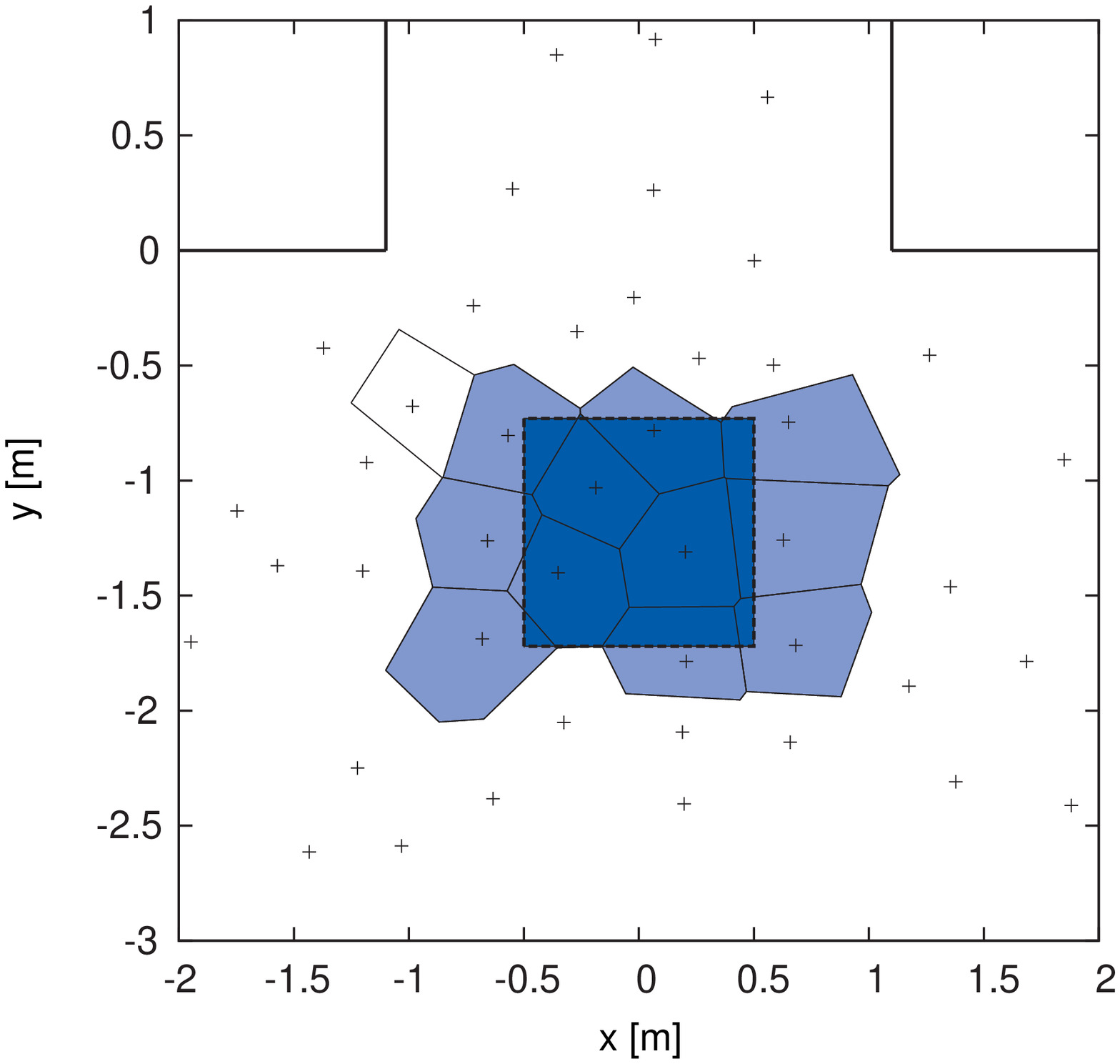}}
\caption{Different definitions of densities in an observation area
  $A$. Definition of $D_s$ is according Eq. \ref{DSTAND}, while $D_V$ and
  $D_{V'}$ is according Eq. \ref{DVORONI} and Eq. \ref{DVORONISTR}
  respectively. The dark blue area contributes to the densities $D_{V'}$ and
  $D_{V}$. \label{DENSDEF}} 
\end{center}
\end{figure}

\begin{figure}[thb]
\begin{center}
    \subfigure[]{\label{Time:xsuba} 
      \includegraphics[width=5.7cm,angle = -90]{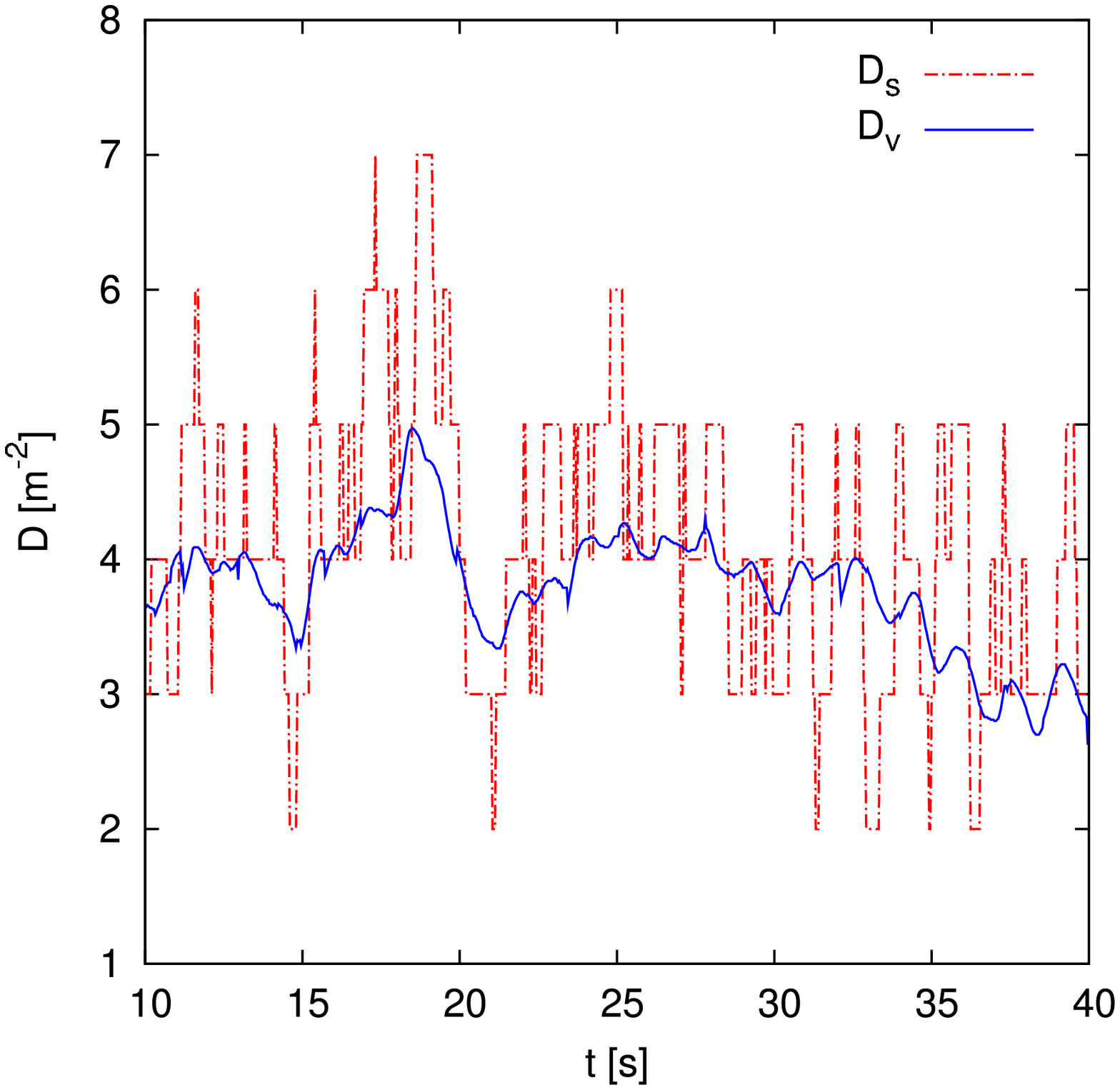}}
    \hspace{20.pt}
    \subfigure[]{\label{Time:xsubb}
    \includegraphics[width=5.7cm,angle = -90]{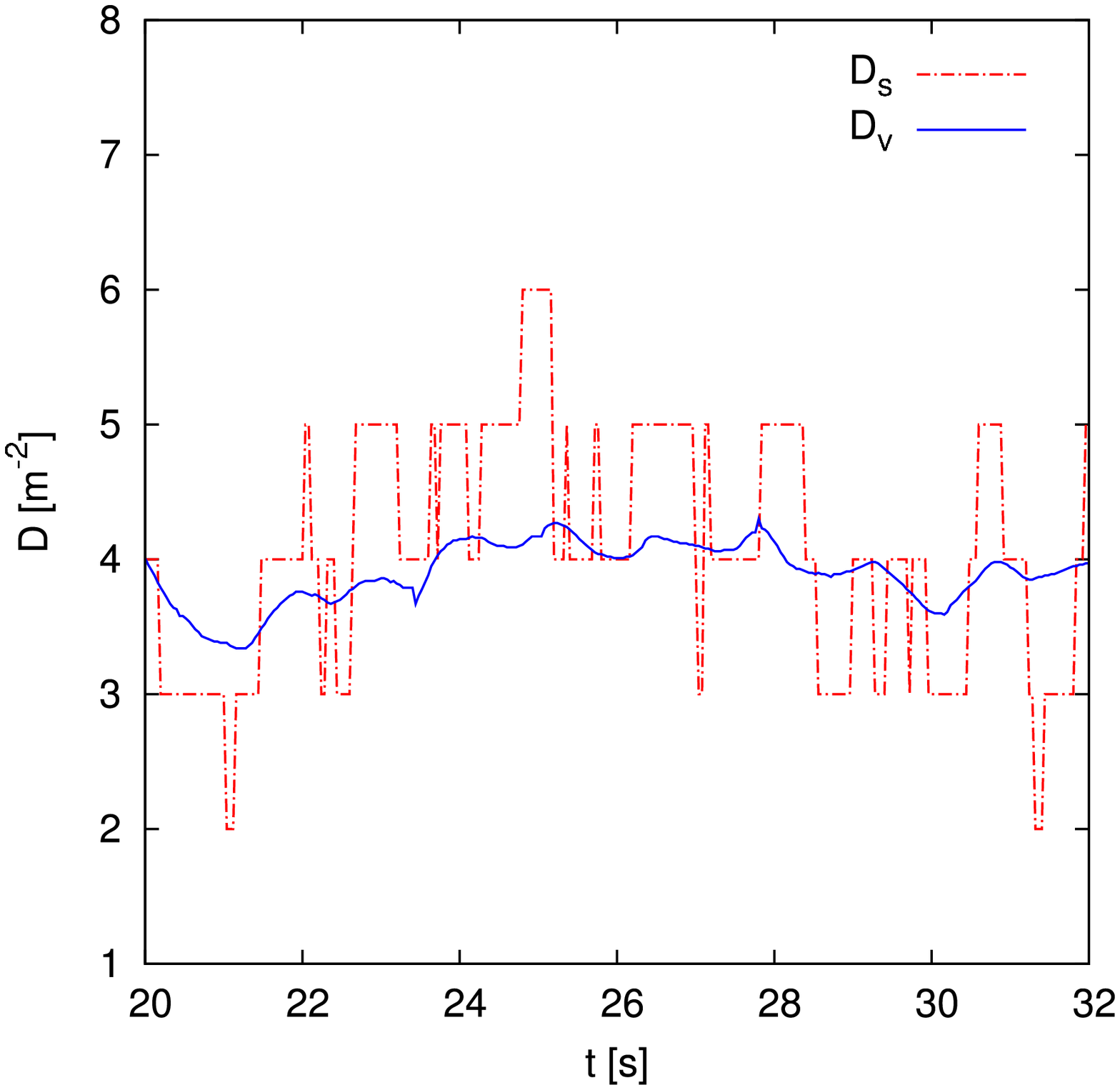}}
    \vspace{20.pt}
    \subfigure[]{\label{Time:xsubc} 
      \includegraphics[width=5.7cm,angle = -90]{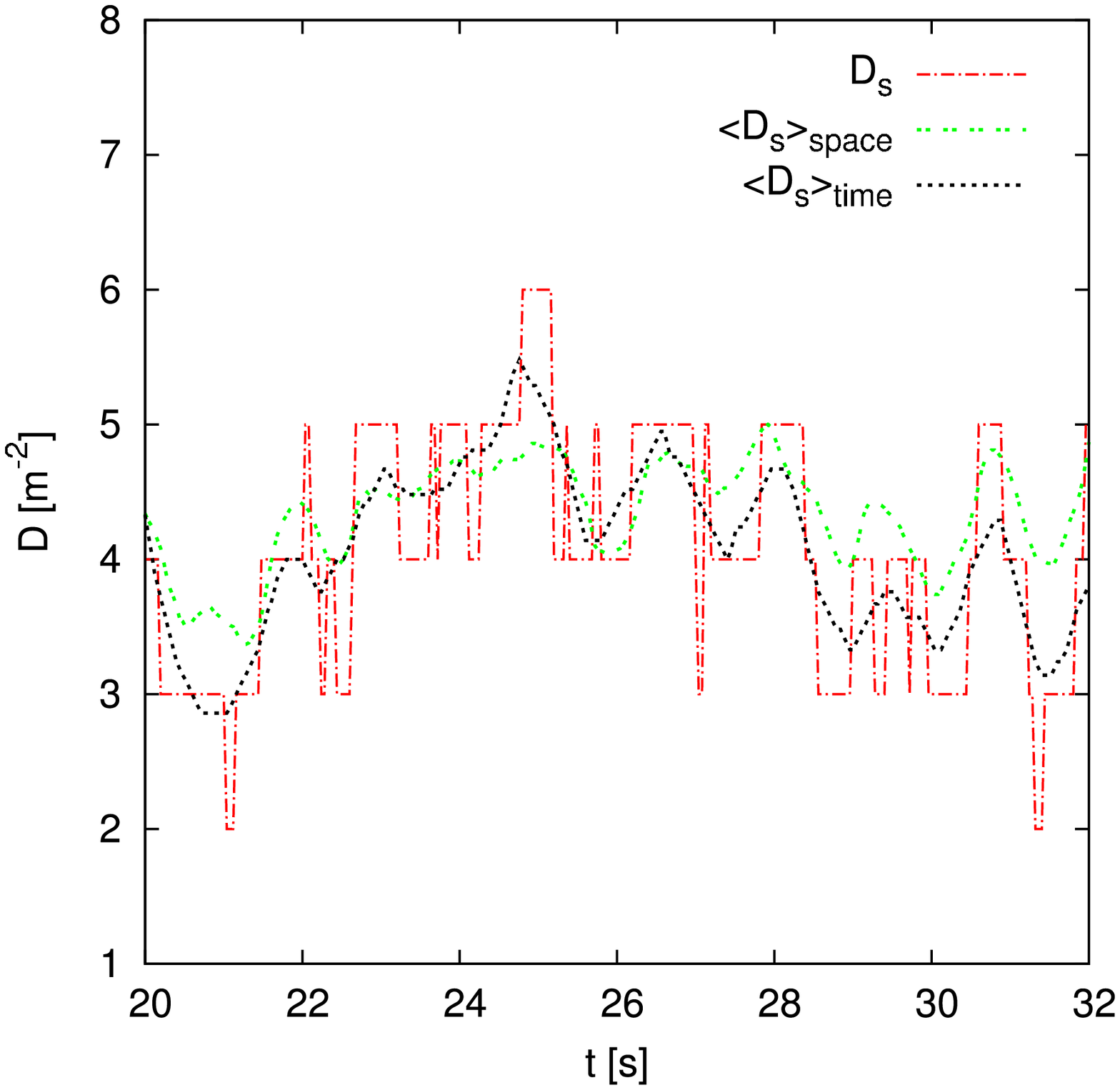}}
    \hspace{20.pt}
    \subfigure[]{\label{Time:xsubd}
      \includegraphics[width=5.7cm,angle = -90]{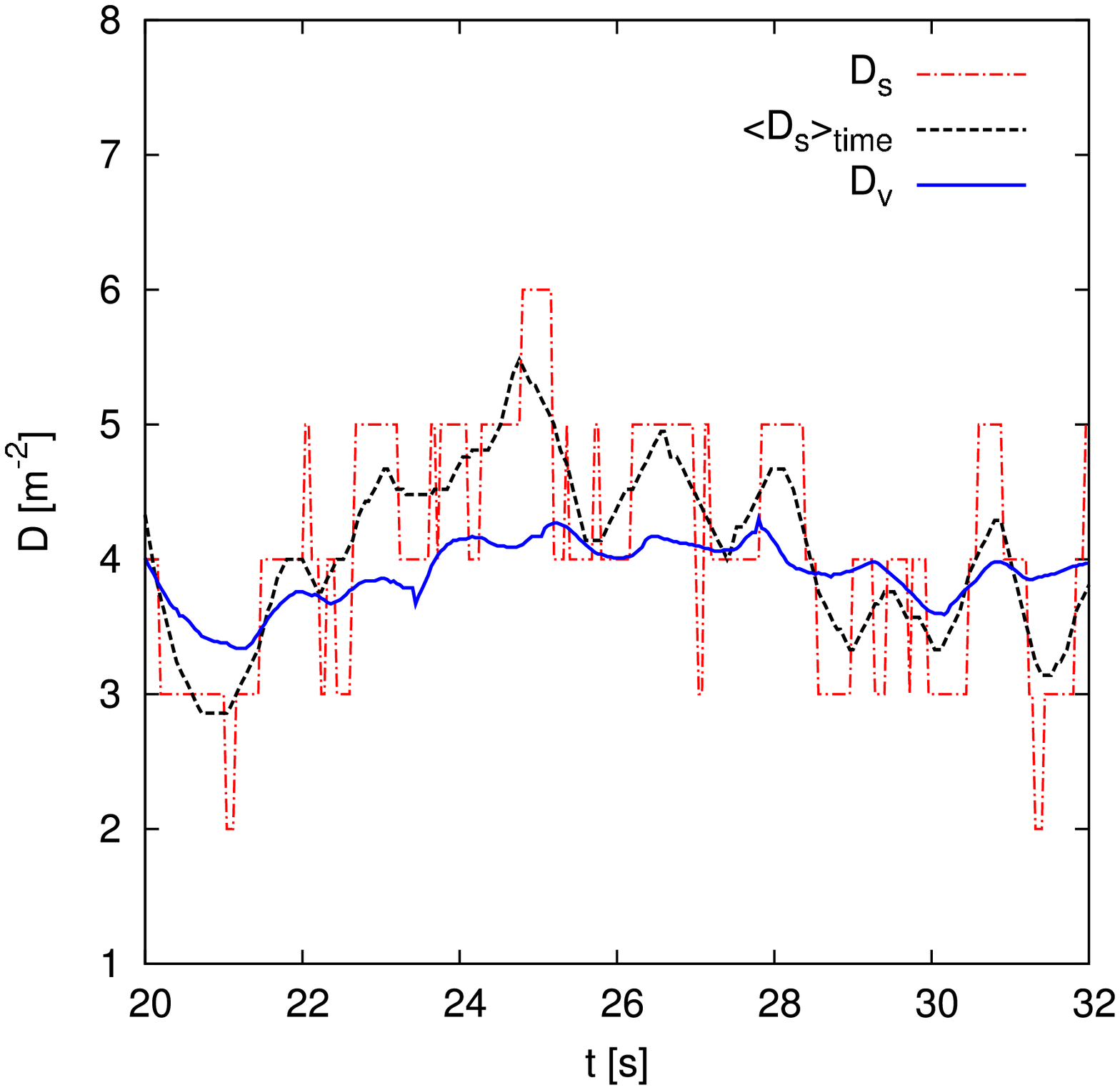}}
\caption{Time sequence of standard density $D_s$ and Voronoi density
  $D_V$. (a): Total run; (b): 'Quasi' stationary state $t=[20,32] s$; (c):
  $D_s$, $\langle D_s\rangle _{space}$ and  $\langle D_s\rangle _{time}$; (d):
  $D_s$, $D_V$ and $\langle D_s\rangle _{time}$.\label{Time}} 
\end{center}
\end{figure}

Fig. \ref{Time} gives a time sequence of $D_s$ and $D_V$. The latter is much
more stable in time, with less than half the standard deviation. While $D_s$
changes repeatedly by about 30\% and back within a second, there is only 
one fast change in $D_V$, and this is less than 10\%. Thus while $D_s$ at a
single time gives no useful information, $D_V$ does.

Fig.  \ref{Time} shows that the overall picture is similar for all definitions
of densities. All densities are measured in the square $A$ of $|A| = 1\;
m^2$, see Fig. \ref{DENSDEF}. $D_s$ obviously  
varies much more than the others, and more than the perception of the situation 
of a human
observer, because entering or exiting of the rectangle by a person changes
$D_s$ considerably, but not the perception. $\langle D_s\rangle _{space}$ was averaged over
rectangles shifted by up to 50cm in each direction, $\langle D_s\rangle _{time}$ is a moving
average over 21 frames ($0.84 s$). $\langle D_s\rangle _{time}$ shows more variation in time
than $\langle D_s\rangle _{space}$, which again varies more than $D_V$. $D_V$ has the
advantage of having full resolution in space and time combined with low
fluctuations. 
  
\begin{table}
\begin{center}
\begin{tabular} {|l|r|r|r|} 
\hline 
\it  Observable $O$ & $\bar{O}$  & $\sigma(O)$  & $TV(O)$ \\ 
           & $[m^{-2}]$  & $[m^{-2}]$ & $[m^{-2}]$ \\ 
\hline
$D_s$               & 4.06 &  0.88 & 47.0 \\
$\langle D_s\rangle _{space}$      & 4.33 &  0.40 & 11.7 \\
$\langle D_s\rangle _{time}$       & 4.07 &  0.62 & 13.7 \\
$D_V$               & 3.90 &  0.23 & 4.7 \\
$D_{V'}$             & 3.83 &  0.29 & 10.6 \\\hline 
\end{tabular}
\end{center}
\caption[dummy]{Mean value $\bar{O}$, standard deviation $\sigma(O)$ and total
  variation $TV(O)=\sum_j |O_{j+1}-O_j|$ for different density definitions for
  results from Fig. \ref{Time}, $t=[20,32]s$. \label{TAB1}}  
\end{table}

For the different densities from Fig. 3 we have the data in Table \ref{TAB1}:
The difference between the density averages for $D_s$ and $D_V$ are within the
limits of the fluctuations, see Table \ref{TAB1}, but it is obvious that the
density distribution is not homogeneous over the entire camera area. The
Voronoi cells carry information from outside the rectangle, where the density
may be different. This may be the reason for part of the differences. \\
Using Voronoi cells, a density distribution $p(\vec{x})$ is attributed to every 
point in space. However, this distribution oscillates with stepping, so for best
results only time averages have to be taken over the time of at least a step, 
or some smoothing of the oscillations is needed (see below). 

\subsection{Measuring velocity}

Given trajectory $\vec{x}_i(t)$ of a person $i$, one standard definitions of
the velocity with $\Delta t$ fixed, but arbitrary is 

\begin{equation}
\vec{v}_{\Delta t,i}(t)=\frac{\vec{x}_i(t+ \Delta t/2)-\vec{x}_i(t- \Delta
  t/2)}{\Delta t}. 
\label{VELSTAND}
\end{equation}

Alternatively, with given entrance and exit time $t_{in},t_{out}$  the
velocity is  

\begin{equation}
\vec{v}_{\Delta x,i}(t)=\frac{\vec{x}_i(t_{out})-\vec{x}_i(t_{in})}{t_{out} -
  t_{in}} \qquad, \qquad  t \in [t_{in},t_{out}] 
\end{equation}

with $s=\|\vec{v}\|$. The average standard speed in an observation area A is
then  

\begin{equation}
 \bar{s}_s(t) = 1/N \, \sum_{\vec{x}_i(t) \in A} \, s_{\Delta t,i}(t) 
\end{equation}

where the sum is taken over all persons that are in A for the entire time
interval $ [t- \Delta t/2 , t+ \Delta t/2 ] $. These definitions seem simple,
but there are two sources of uncertainty. The velocity of an extended object
is generally (and reasonably) defined as the velocity of its center of mass,
and for pedestrians that is hard to detect. Moreover, pedestrians can and do
change shape while walking. So the simple approach works only for distances
long enough to make errors from shape changes and in  placing the supposed
center of mass unimportant. A second problem comes from the fact that velocity
is a vector, and the movement of people is not straight. Thus the average of
the local speeds will be bigger than the value of distance per time for longer 
distances. Notably head tracking gives tracks that can be decomposed into a
fairly uniform principal movement and a local swaying superimposed. The
swaying shows the steps, it varies between individuals and is larger at low
speeds. Of course, the head sways more than the center of mass, and it can do
some independent movements, but these can dominate only at very low
speeds. For most models of pedestrian movement, only the principal movement is
of interest. The swaying movement of shoulders may be important in estimating
the necessary width for staircases and corridors to allow overtaking.

The separation of principal movement from swaying movement could be done by
Fourier analysis, but that requires a trajectory many steps long. A way to do
it locally is to detect positions of identical phase of the movement and
interpolate them. As long as there is appreciable forward motion ($>
0.3\;m/s$), the mode of movement is the swinging of the legs in the direction
of movement with approximately their free pendulum frequency (1.5Hz-2Hz). 
In this mode there is a regular sequence of points of maximum (positive)
curvature, minimum (negative) curvature, and zero curvature, which correspond
to the times of setting down the right foot, the left foot, and having one
foot on the ground while the other just passes the standing leg in forward
motion, which are the points we use. These points are easy to detect. Below
that speed, the mode of stepping changes to the whole body swinging right and
left with a frequency smaller than 1Hz and only a small forward component, and
there may be multiple points with zero curvature $d^2\vec{x}_i=0$ within one step. 
However, typically the positions of setting down a foot give a dominant extremum of
curvature, while in the part in between the curvature will be close to zero
with more than one zero per step. In this case, we take the middle between the
maximum and the minimum curvature point as interpolation point. This has been
possible down to speeds of 5 cm/s. Below that, steps can only be guessed, they
can not be detected reliably.  

The speed of the principle movement $s_{p,i}=\|\vec{v}_{p,i}\|$ is calculated by
\begin{equation}
\vec{v}_{p,i}=\frac{\vec{x}_i(t_2)-\vec{x}_i(t_1)}{t_2 - t_1} \quad \quad
\mbox{with} \quad \quad \mbox{curvature}(\vec{x}_j(t_j))=0, \quad j=1,2.
\label{VELSTEP}
\end{equation}

\begin{figure}[thb]
  \begin{center}
    \includegraphics[width=7cm,angle = -90]{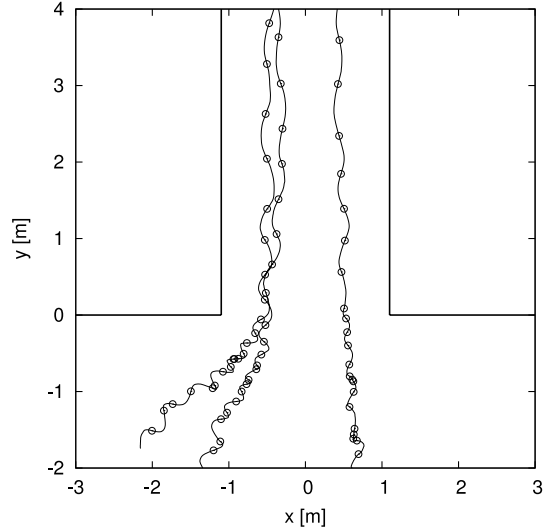}
\caption{Trajectories of three persons walking through bottleneck. Circles
  denote interpolation points for a smoothed trajectory. \label{Traj}}     
  \end{center}
\end{figure}

Interpolating these points now eliminates most of the swaying movement and
gives a good approximation to the movement of the center of mass. The
requirement of identical phase asks for taking only every other zero curvature
position, but for persons with symmetric gait, taking every zero curvature
position gives a better resolution with only marginally more swaying. For
analysis, we will take this as curve of principal movement.

Similarly, the velocity vector will be obtained by computing the difference
quotient of position and time between zero curvature positions, and attached
to the intermediate time.
 
\begin{figure}[thb]
  \begin{center}
    \includegraphics[width=7cm,angle = -90]{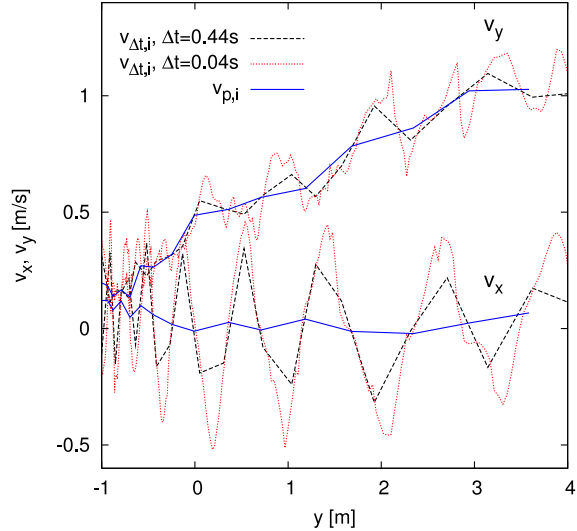}
\caption{Speed in x and y versus time for part of the middle trajectory of
  Fig. \ref{Traj}; dotted(red): $\Delta t = 0.04 s$; dashed(black): $\Delta t
  = 0.44 s$ according Eq. \ref{VELSTAND}; solid(blue): as described in
  Eq. \ref{VELSTEP}. \label{Speed}}    
  \end{center}
\end{figure}
    
\subsection{Measuring flow}

Standard measurement of flow $F_s$ through a door or across a line is done
similar to density measurements, by counting heads passing within a time 
interval. This suffers from the same problems as the standard density
measurement, namely large scatter and low time resolution. Using the Voronoi
cells to obtain fractional counts (half a person has passed if half of the
Voronoi cell has passed) gives a much smoother Voronoi flow $F_V$. This still
does not allow a useful passing to the limit of $\Delta t = 0$, but the moving
average over about half the average time difference between persons passing
gives a sufficiently smooth result. 

\begin{figure}[thb]
  \begin{center}
    \includegraphics[width=6cm,angle = -90]{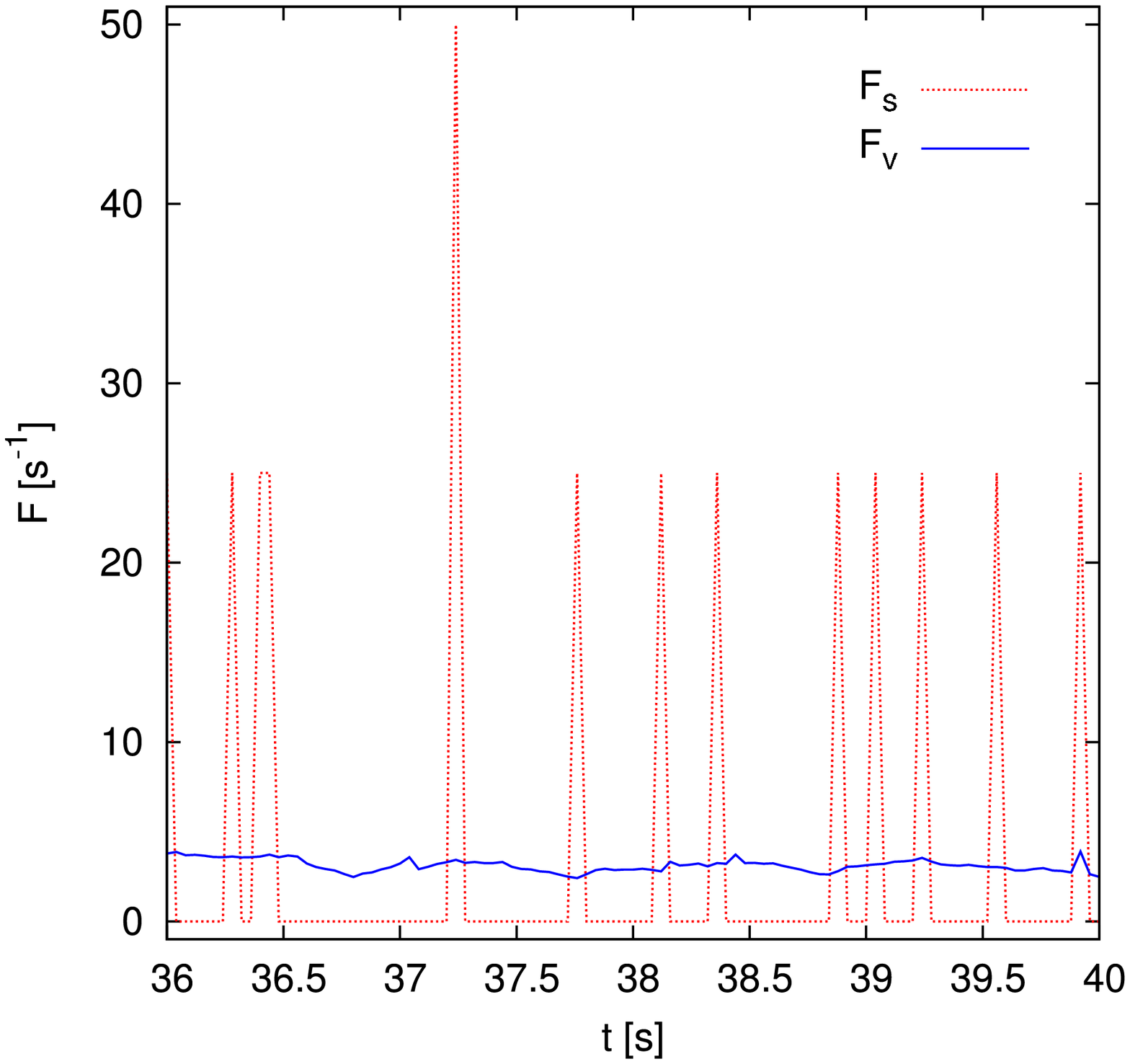}
\qquad
    \includegraphics[width=6cm,angle = -90]{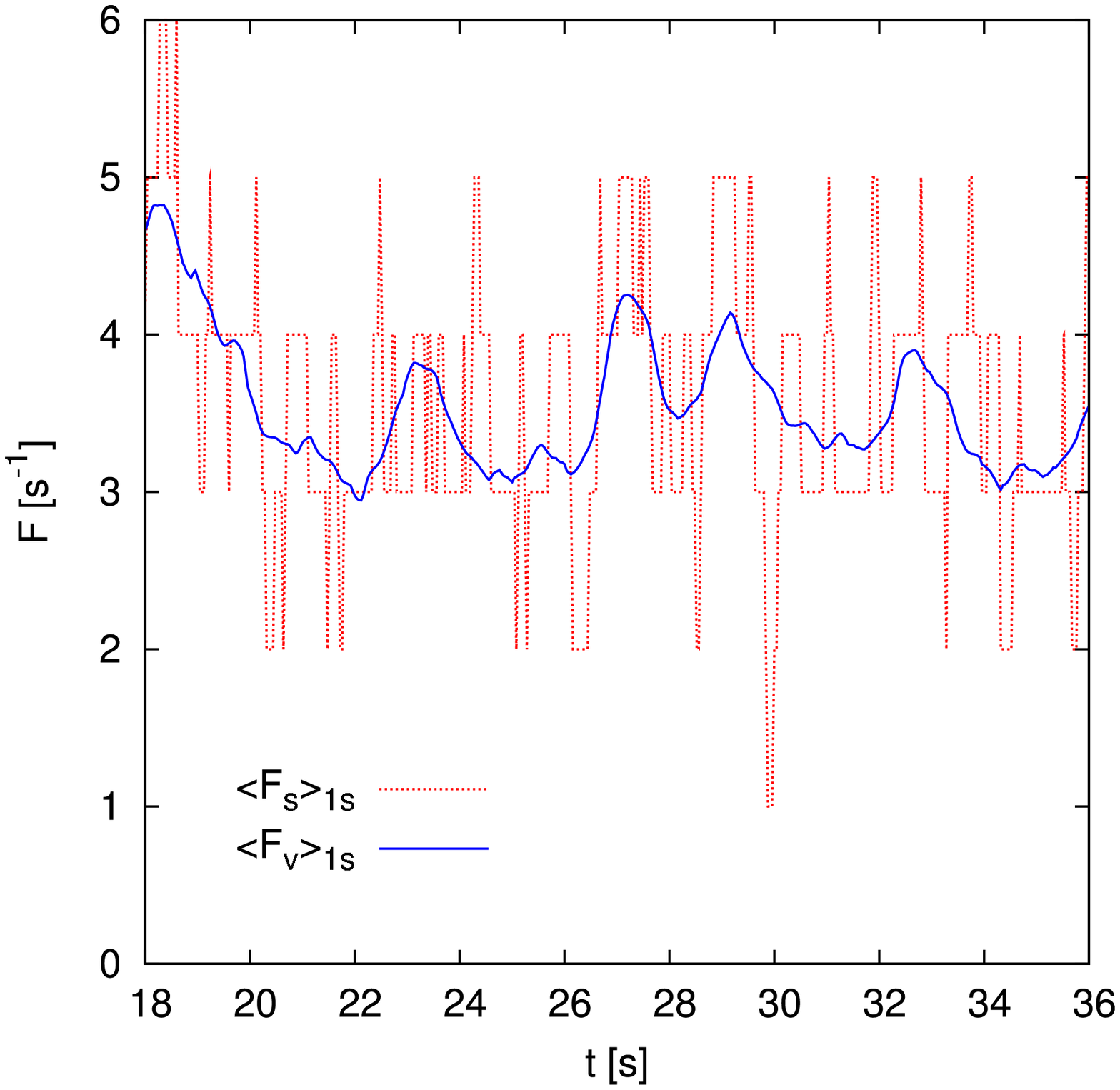}
\caption{Flux through bottleneck with standard method $F_s$ and Voronoi cell
  based method $F_v$, left: frame by frame  right: 1 s moving
  average. \label{Flux_movav}}    
  \end{center}
\end{figure}
\hspace{0.1mm}

\begin{figure}[thb]
  \begin{center}
    \includegraphics[width=7cm,angle = -90]{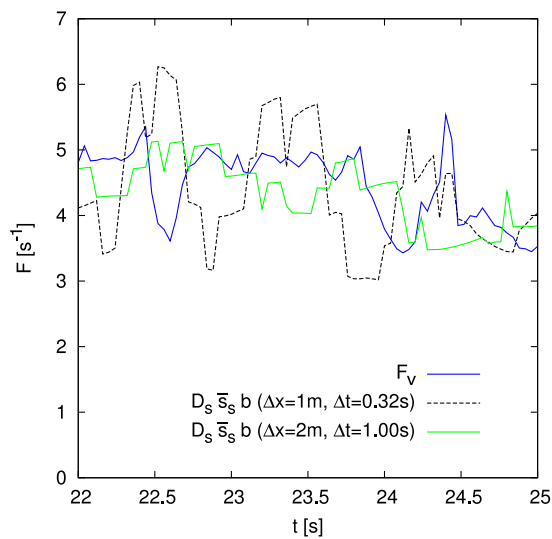}
\caption{Flux at bottleneck. Voronoi method $F_v$ and flow computed by
  $F=D_s\,\bar{s}_s\, b$ where b is the width of the bottleneck.\label{Fl_co} }  
  \end{center}
\end{figure}

Fig. \ref{Fl_co} compares the product of averaged standard speed and density,
$\bar{s}_s(t) \, D_s (t)$, with the direct measurement of the flow as Voronoi
density passing the middle line. The density was determined from a rectangle
symmetric to the line for the flow of width $\Delta x$ and the averaged speed
of the persons inside this rectangle from a symmetric time difference of
$\Delta t$. The two product evaluations  agree reasonably well with the
Voronoi flow, but in spite of the fact that the measurements average over time
and space they show much faster variations in time, and the total flow they
calculate is somewhat less than the number of persons passing. For $\Delta
x=1m$, $\Delta t=0.32s$ one person of 180 is missing in the integration, for
$\Delta x=2m,\; \Delta t=1s$ it is even 7 persons missing. Depending on the
purpose of the measurements, this may be a serious problem.

\section{Prospects of analysis}

\subsection{Fundamental diagram at bottlenecks}

\begin{figure}[thb]
  \begin{center}
    \includegraphics[width=8cm,angle = -90]{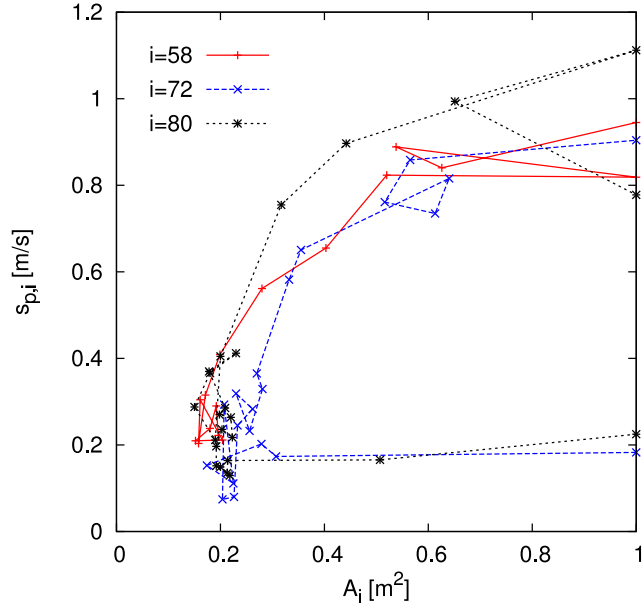}
\caption{Personal space versus speed for three persons,
  comp. Fig. \ref{Traj} \label{sp_vel}}     
  \end{center}
\end{figure}

\begin{figure}[thb]
  \begin{center}
    \includegraphics[width=6cm,angle = -90]{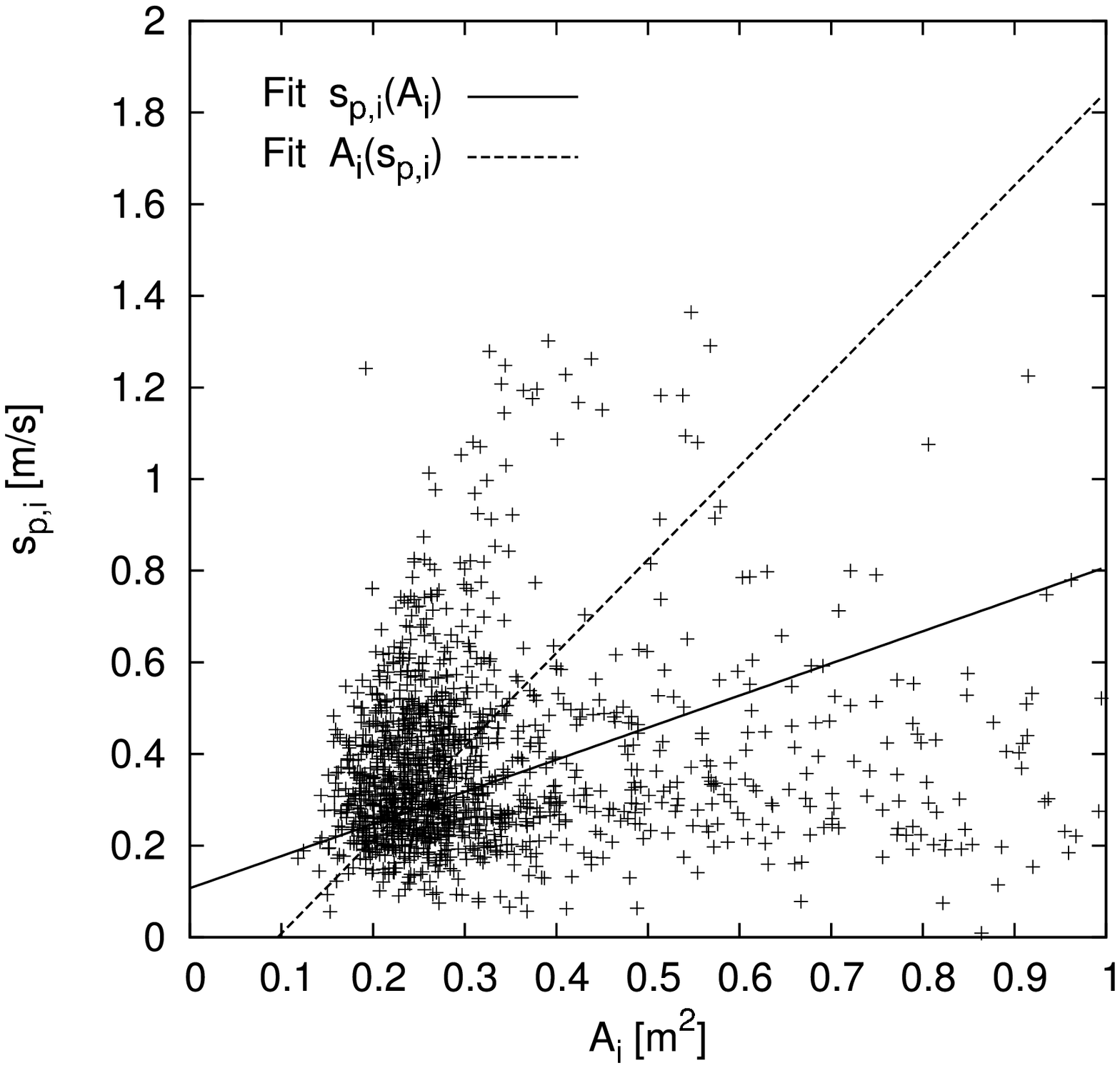}
\qquad
    \includegraphics[width=6cm,angle = -90]{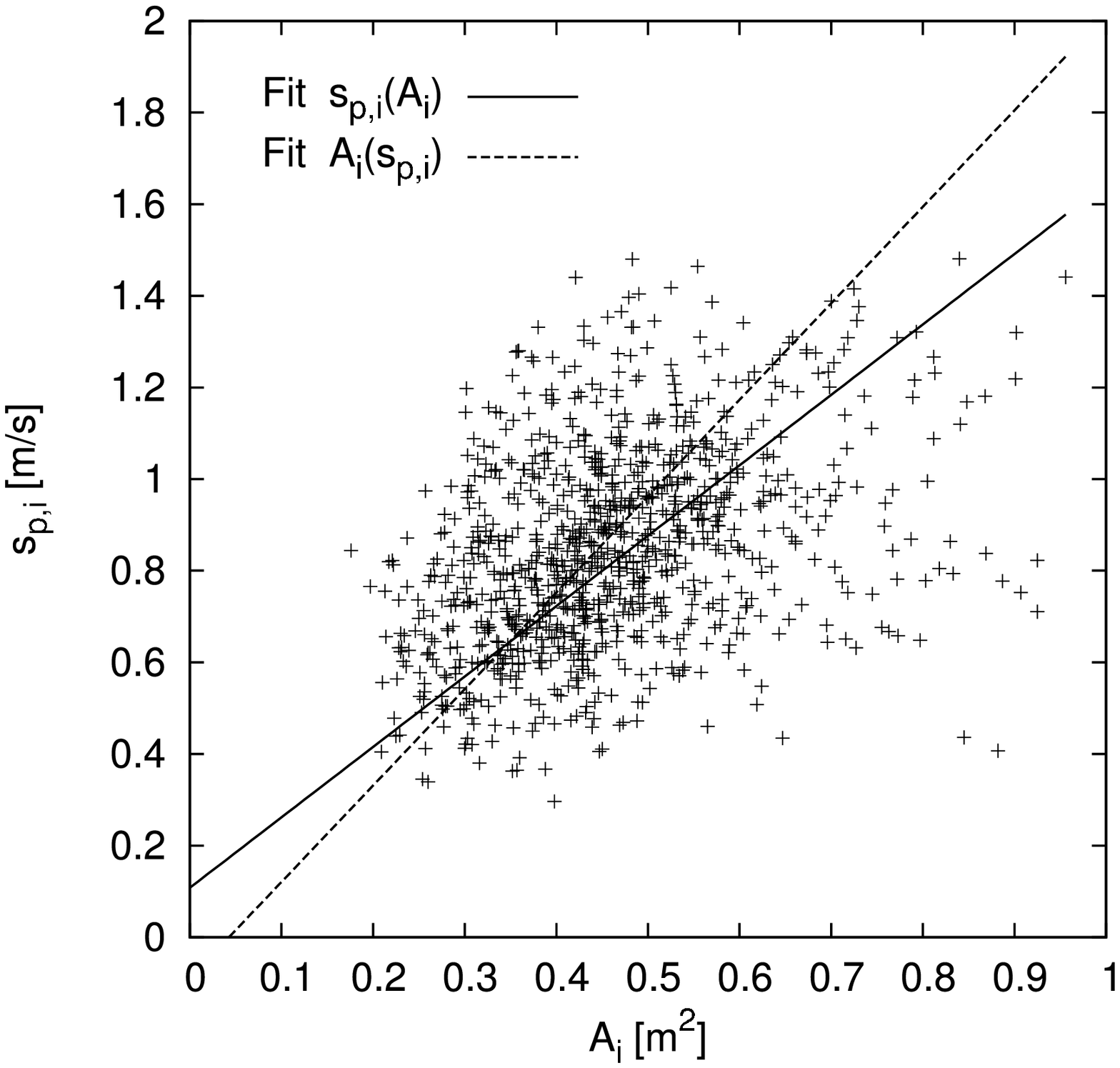}
\caption{Personal space versus speed, left: in front of bottleneck right: in
  bottleneck}  \label{sp_v_gen} 
  \end{center}
\end{figure}

The high resolution measurement of speed and density allows to follow
individuals on their way through some obstacles and look what combinations
of speed and density they have. In Fig. \ref{sp_vel} the persons start 
in front of the bottleneck with low speed, no. 72 and 80 with large space
while no. 58 is already in the jamming area. They pass the
congestion area and pick up speed inside the bottleneck. It also allows
correlating momentary speed and personal space, as well as comparing personal
space and speed in general for whole groups of persons. Fig. \ref{sp_vel} and
\ref{sp_v_gen} show a substantial difference in this relation before and
inside the bottleneck, indicating that the individual speed depends more on the
expectation of the future (walking into or out of high density regions) than
on the present situation. The high resolution makes it also possible to
analyze how the accelerations observed are related to changes in the space
available, and to the space of people in front. On the resolution presented 
here it becomes clear that the correlation of speed and available space in an 
instationary situation differs considerably from that in the stationary 
situation described by the fundamental diagram.

\subsection{Fundamental diagram for single file movement}

The most important relation for any model (and for much of the analysis of
pedestrian movement) is the so-called fundamental diagram, which can be given
either as relation speed versus density or flow versus density. Using the
improved methods of measurement, the quality of the resulting diagram is
greatly enhanced.  

\begin{figure}[thb]
  \begin{center}
    \includegraphics[width=6cm,angle = -90]{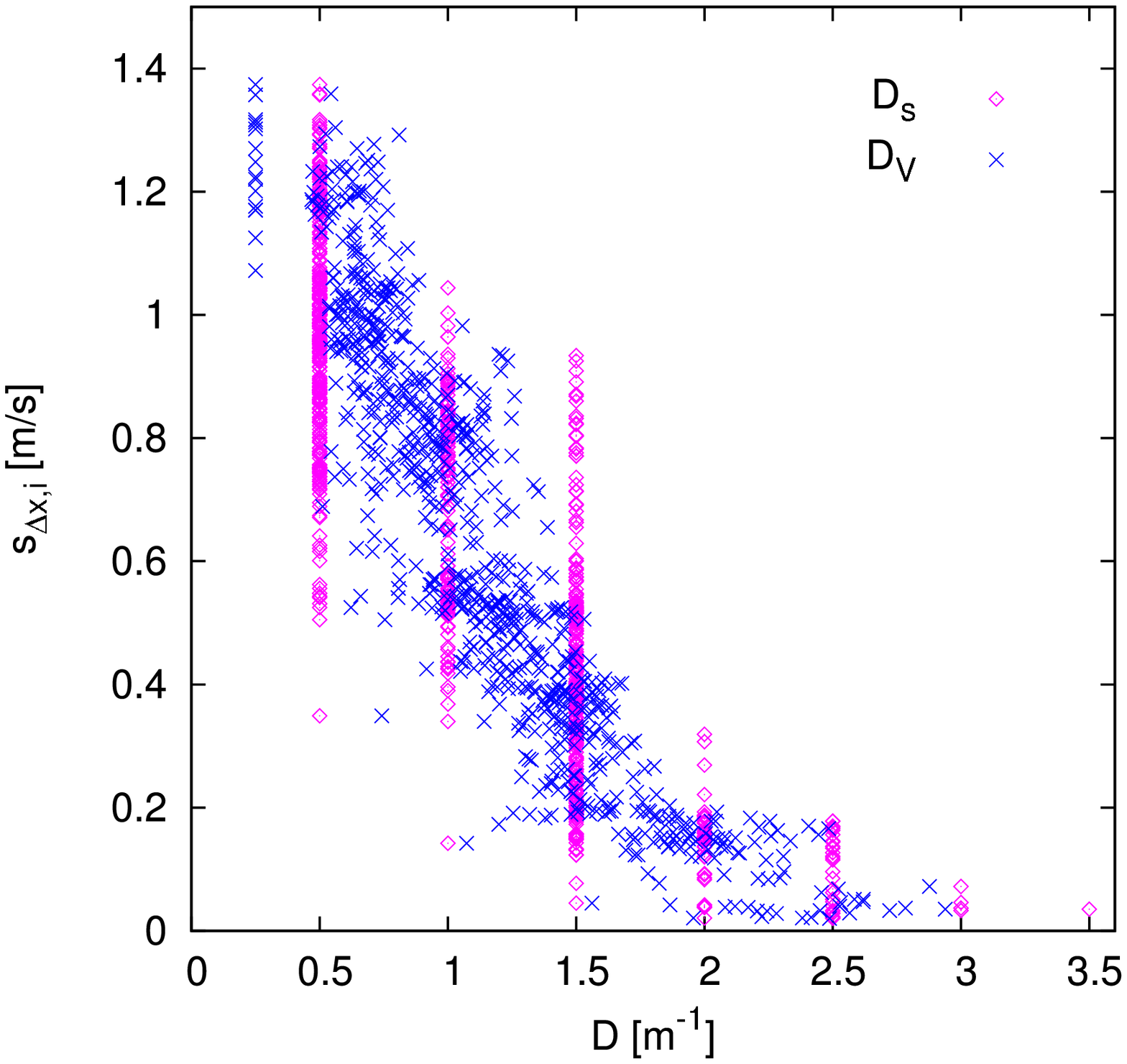}
\qquad
    \includegraphics[width=6cm,angle = -90]{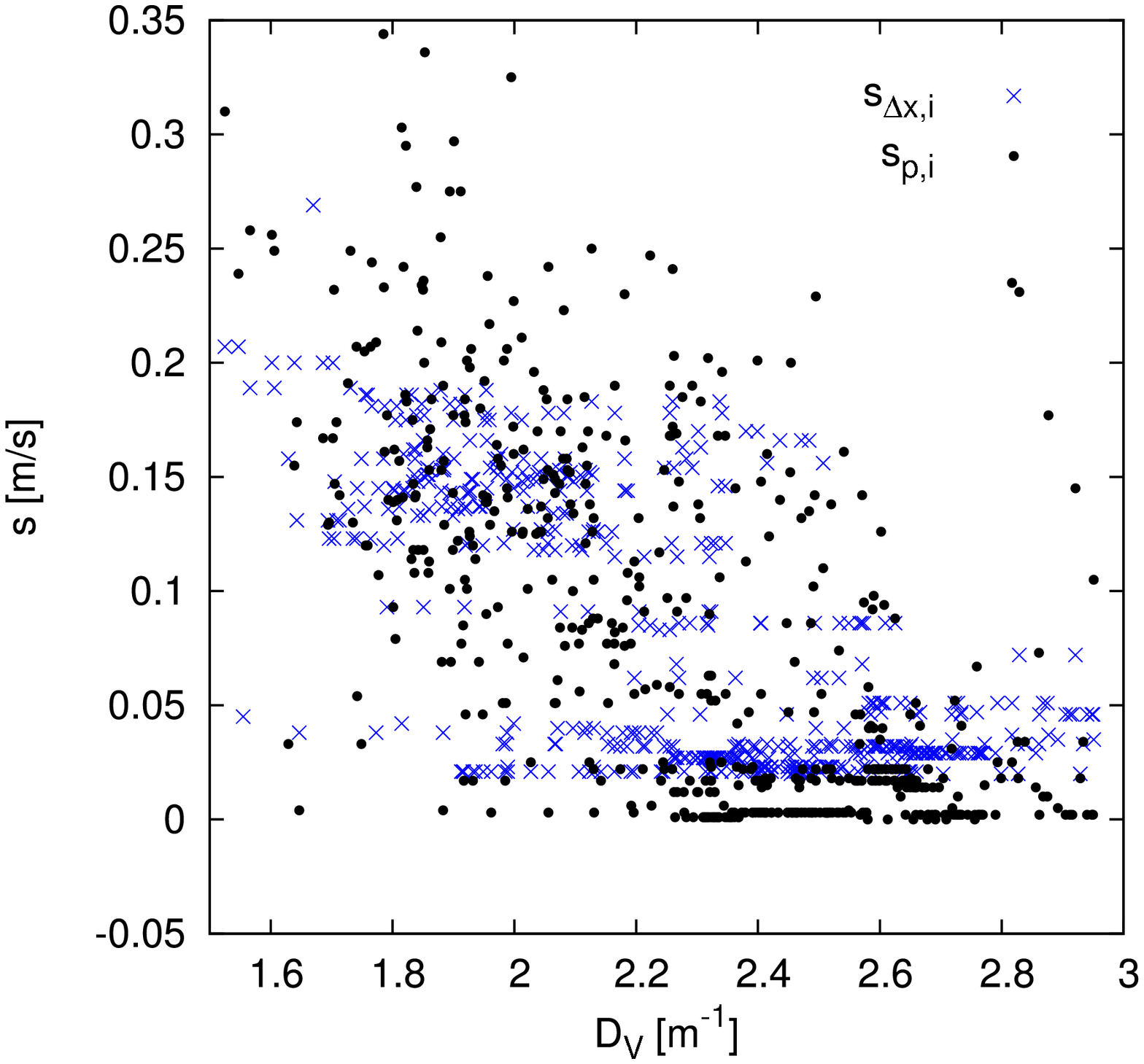}
\caption{Speed versus density for analysis with different methods. The
  trajectories are from an experiment of single file movement, see
  \cite{Seyfried2009a}. Left: sampled when a person crosses a certain
  line. Right: sampled every second, high densities only.  Diamonds(red):
  $s_{\Delta x,i}$ with $\Delta x = 2m$ versus density $D_s$. Crosses(blue):
  same speed, density $D_V$. Points(black): speed from step detection
  $s_{p,i}$, density $D_V$. \label{FD}}     
  \end{center}
\end{figure}

As an example Fig.(\ref{FD}) shows the analysis of a series of  measurements
of the fundamental diagram for single file movement performed with
different numbers of persons (14 to 70) in the walking area, which covers all
densities of interest from low (almost free walking) to jamming density (near standstill). With
the high density, stop-and-go waves developed, see Fig. 5 in
\cite{Seyfried2009a}. For the one dimensional case with position $x_i$ of
pedestrian $i$ the calculation of the Voronoi density distribution reduces to 

\begin{equation}
 p_i(x) =  \left\{\begin{array}{r@{\quad:\quad}l}
\frac{2}{|x_{i+1}-x_{i-1}|} & x \in [\frac{x_{i-1}+x_i}{2},\frac{x_{i+1}+x_i}{2}] \\
0 & \mbox{otherwise}
\end{array} \right.  
\end{equation}

The definition of density $D_V$ is according to
Eq. \ref{DVORONI}. Fig. (\ref{FD}) (left) shows that use of $D_V$ greatly
enhances the quality of the diagram,  due to the fine density resolution. Use
of the 
standard density contracts the band of s versus D onto a few vertical lines 
which are much longer than the width of this band and thus reduces the
precision. In Fig.(\ref{FD}), (right) one can see that with the standard speed
$s_{\Delta x,i}$ the lowest speed is about 0.02 m/s, while $s_{p,i}$ gives values of
zero. Actually, some people were standing in one position for about 30s - the
duration of a stop phase. For higher speeds, there is little difference
between $s_{\Delta x,i}$ and $s_{p,i}$. The combination of  $D_V$ and $s_{p,i}$ gives the best
diagram for the full scale of densities.

\section{Concluding remarks}
 
The combination of modern video equipment with new methods for extracting
relevant data allows an unprecedented depth of analysis of pedestrian
behavior. The method for determining density is based on the concept of a
Voronoi cell as personal space of a pedestrian and allows a resolution down to
individual level. The concept of determining velocities from difference
quotients of positions with identical phase of the stepwise movement gets the
resolution down to a single step. This level of resolution allows
mathematical combination of data that are not valid for large scale averages. 
Moreover they are able to resolve stop and go waves and allow a analysis of
instationary processes on a microscopic level.

\section*{Acknowledgment}
The experiments are supported by the DFG under grant KL 1873/1-1 and SE
1789/1-1. We thank M. Boltes for his support in preparation of videos for
analysis.   

\bibliographystyle{plain}

\end{document}